\documentclass[3p,times]{elsarticle}

\usepackage{ecrc}

\volume{00}

\firstpage{1}

\journalname{Procedia Computer Science}

\runauth{}

\jid{procs}

\jnltitlelogo{Procedia Computer Science}

\CopyrightLine{2011}{Published by Elsevier Ltd.}

\usepackage{amssymb}
\usepackage{amsmath}

\usepackage[figuresright]{rotating}

\begin{document}

\begin{frontmatter}



\dochead{}

\title{Hole Detection and Shape-Free  Representation and Double Landmarks Based Geographic Routing in Wireless Sensor Networks}


\author[label1]{Jianjun Yang}
 \address[label1]{Department of Computer Science and Information Systems, University of North Georgia, Oakwood, GA. Email: jianjun.yang@ung.edu}
\author[label2]{Zongming Fei}
\address[label2]{Department of Computer Science, University of Kentucky, Lexington,KY. Email: fei@netlab.uky.edu}
\author[label3]{Ju Shen}
\address[label3]{Department of Computer Science, University of Dayton, Dayton, OH. Email: jshen1@udayton.edu}

\address{}

\begin{abstract}
In wireless sensor networks, an important issue of Geographic Routing is ``local minimum" problem, which
is caused by ``hole" that blocks the greedy forwarding process. Existing geographic
routing algorithms use perimeter routing strategies to find a long detour
path when such a situation occurs. To avoid the long detour path, recent research
focuses on detecting the hole in advance, then the nodes located on the boundary of the hole
advertise the hole information to the nodes near the hole. Hence the long detour
path can be avoided in future routing. We propose a heuristic hole detecting algorithm which
 identifies the hole easily and quickly and then propose a representation of hole no matter what the shape of the hole is. In addition, we quantitatively figure out the areas
in the vicinity of the hole that need to be announced the hole information. With such information,
a new routing scheme with two landmarks was developed.
Simulation results illustrate
that our approach can achieve better performance in terms of the average length and number of hops in routing
paths. Simulation also shows that our approach introduces very small computational complexity.
\end{abstract}


\end{frontmatter}


\section{Introduction}
\label{Int}
Wireless  Sensor Networks(WMSNs) have emerged
as one of the key technologies for wireless communications.
They are undergoing rapid development and have inspired
numerous applications\cite{jshen1}\cite{jshen2}\cite{jshen3}\cite{vehicle1}\cite{channel1} because of their advantages. 
A Wireless Sensor Network consists of a collection of wireless
communication nodes. Two nodes within a certain distance
of each other can communicate directly. However, if a source
node intends to send packets to a destination outside of its
transmission range, it will depend on other nodes to relay the
packets.
Many routing protocols (e.g., DSDV \cite{dsdv}, AODV \cite{aodv})
have been proposed 
to find the
path from the source to the destination. The main research issue with
these routing schemes is the scalability because most of them
have to use flooding to find routing paths.

When the location information for nodes is available (either
through GPS or using virtual coordinates \cite{oldrouting1}), routing in sensor networks  can be much more efficient. Geographic routing
exploits the location information and makes the routing in
sensor networks scalable. The source node first acquires the
location of the destination node it intends to communicate with,
then forwards the packet to its neighbor closest to the destination. This process is repeated until the packet reaches the
destination. A path is found via a series of independent local
decisions rather than flooding. However, geographic routing
suffers from the so-called local minimum phenomenon, in
which a packet may get stuck at a node that fails to find a
closer neighbor to the destination, even though there is a path
from the source to destination in the network. This typically
happens when there is a void area (or hole) that has no active
nodes. In wireless sensor network, the holes are caused by
various reasons \cite{holereason}. For instance, the malicious nodes can jam
the communication to form Jamming Holes. If the signal of
nodes is not long enough to cover everywhere in the network
plane, the Coverage Holes may exist. Moreover, Routing Holes
can be formed either due to voids in node deployment or
because of failure of nodes due to various reasons such as
malfunctioning, or battery depletion.

To deal with the local minimum problem, Karp and Kung
proposed the Greedy Perimeter Stateless Routing (GPSR)
protocol, which guarantees the delivery of the packet if a path
exists \cite{GPSR}. When a packet is stuck at a node, the protocol will
route the packet around the faces of the graph to get out of
the local minimum. Several approaches were proposed that are
originated from the face routing. Although they can find the
available routing paths, they often cause the long detour paths.

Current research is focused on developing algorithms to overcome the local minimum issue in geographic routing by finding holes prior to packet forwarding towards the holes.
Scholars may use particular approaches to define and find holes in some real work applications.
For instance, in a sensor network that monitors temperature in a region, if we let a sensor node mark itself as unavailable once its
local temperature exceeds a threshold, then the boundary of a hole can probably be determined based on the temperatures of the nodes. Such a hole is represented as a polygon that encloses all the sensors with local temperatures higher than the threshold. Unfortunately, these algorithms are time or space consuming. Moreover, the representation of a hole is too complicated.
Most recent work tries to detect a hole and the nodes located on the hole's boundary
in advance ~\cite{fang31}~\cite{xing45}. The nodes on the boundary further advertise the hole information to some other nodes. In this way, the future routing path can be adaptive in the presence of the hole. In this chapter, we introduce an algorithm
of Shape-Free Hole Representation and Double Landmarks Based
Geographic Routing
for wireless sensor networks. It focuses on defining and detecting holes in a wireless sensor network, representing holes and building routes around the holes. It is a heuristic algorithm aimed to detect a hole quickly and easily.
The hole can be identified by a constant time complexity calculation.
In addition, we provide a very concise format to represent a hole by representing a hole as a segment.
Moreover, we develop an approach to make part of the nodes located on the hole's boundary announce to the nodes
in the vicinity of the hole.
We further found the best trade-off between the overhead of hole information announcement and the benefit for future routing.





\section{Related Work}
\label{related}

The first geographic routing protocol is based on simple greedy forwarding. In this approach,
each node forwards packets to one of its neighbors who is closest to the destination node until
the packets arrive the destination. This scheme is efficient. However, it fails due to the ``local minimum problem".

To mitigate ``local minimum problem", compass routing~\cite{compass} was proposed as the first face routing,
in which the packet is forwarded along the face until greedy is workable in a node. However,
compass routing cannot guarantee packet delivery in all geographic networks. Several routing
algorithms in face routing family have been developed. By combining greedy and face routing, Karp and Kung proposed the
Greedy Perimeter Stateless Routing (GPSR) algorithm~\cite{GPSR}.
It consists of the greedy forwarding mode and the perimeter forwarding mode, which is applied in the regions where
the greedy forwarding does not work. An enhanced algorithm, called Adaptive Face Routing (AFR)~\cite{kuhn}, uses an eclipse
to restrict the search area during routing so that in the worst
case, the total routing cost is no worse than a constant factor
of the cost for the optimal route. The latest addition to
the face routing family is Path Vector Face Routing(GPVFR)~\cite{leong}, which improves
routing efficiency by exploiting local face information.
The protocols in face routing family can avoid the hole. However, they often cause long detour path.

Two  routing algorithms were proposed  to avoid long detour path caused by hole. One is ITGR~\cite{ITGR}.
The source determines destination areas which are shaded by the holes based on previous
forwarding experience. The novelty of the approach is that a
single forwarding path can be used to determine an area that
may cover many destination nodes. An efficient
method is designed for the source to find out whether a destination
node belongs to a shaded area. The source then selects an
intermediate node as the tentative target and greedily forwards
packets to it to avoid the long detour. Finally the intermediate
target forwards the packet to the destination by greedy routing. The second is HDAR~\cite{YangHDAR}.
A heuristic
algorithm is designed to detect a hole quickly and easily. And the hole can
be identified only by  calculation with constant time complexity. Then
a concise representation of the hole is devised. A hole is recorded as a
segment. Moreover,  an approach that let a subset of
the nodes located on the hole's boundary announce the hole
information to the nodes in the vicinity is developed.

A new idea~\cite{fang31} was proposed recently, which is to detect the hole in advance, then the nodes located
on the hole advertise the hole information to other nodes. The hole information will benefit nodes who receive it for their future routing.
It defined a hole to be a simple region enclosed by a polygon
cycle which contains all the nodes where local minimum can appear. It brought forth the ``get stuck" concept and
proposed the hole detection mechanism that
once a packet  following geographic greedy forwarding gets stuck at a node, the node must be on the boundary of a hole.
Also related is HAGR~\cite{xing45}. HAGR investigated the
nodes incident to a close loop in a geographical graph. For a vertex $u$, if the
angle between two adjacent edges with respect to this vertex is larger than an angle threshold,
then vertex $u$ considers it is located on a potential hole. To further determine if it is located on a hole,
$u$ calculates the diameter of the loop. It locates the bisector that equally splits the angle and uses
it as a reference line. Then node $u$ finds out the leftmost node and the rightmost node furthest from the
bisector. The distance between the leftmost node and the rightmost node is the diameter of the hole.
If the diameter is greater than the diameter threshold and the angle is bigger than the angle threshold,
$u$ is regarded as sitting on a hole. Once a node is detected on  a hole, it advertises the hole information
to its neighbors. Upon receiving the hole information, its neighbor recalculates the angle and diameter
based on its location. If both of them are bigger than their thresholds, then the neighbor considers it is
on a hole and it continues to advertises the hole information, otherwise it stops advertisement.
Base on the hole detecting, HAGR divides the network plane into three regions, and the nodes in different
regions conduct different forwarding strategies. The idea of HAGR is novel. However, the hole detecting
approach is time-consuming since a node has to calculate the values of two metrics. And the hole advertisement is
expensive because once a node receives the hole information, it has to recalculate two values and
compare them with their corresponding thresholds. In addition,  the diameter threshold is an absolute value and it has to be adjusted according to
the nodes' transmission range or the network deployment, otherwise false negative or false positive may occur.
Moreover, the forwarding strategies are too complicated.


\section{Hole Detecting Algorithm}

\subsection{Metric to Determine a Hole}
 We call our Hole Detection and Double Landmarks Based Geographic Routing algorithm HDDL. 
In HDDL, a node $p$ begins to detect whether it is located on the boundary of a hole only if the angle between its two
adjacent edges is greater than 120 degrees~\cite{fang31}.
$p$ initiates a probe message which includes its location. $p$ sends the message to its
leftmost node with respect to the angle. The leftmost node can be defined as follows. $p$ faces the area formed by the two rays of
this angle, and uses the angle's bisector line to conduct counter-clockwise sweeping. The leftmost node is the first one that is met
by the sweeping line. Upon receiving the probe message, $p$'s leftmost neighbor node writes
its location into the message and passes it to its leftmost neighbor. The probe message will finally come back
to node $p$ from $p$'s rightmost neighbor with respect to the initial investigated angle~\cite{compass}~\cite{fang31}, where the right most neighbor is defined in the similar
way as the leftmost node.
When the probe message circulates,
it collects the locations of the nodes on its way. So node $p$ knows all the nodes' locations on the way.

$p$ then begins to investigate the nodes by traveling clockwise from node to node. For each node on the way, $p$ computes the length of their probe path $length\_pro()$
and their Euclidean distance $dist\_euc()$. For a node $x$, $length\_pro(p, x)/dist\_euc(p, x)$ is defined as hole detection ratio from $p$
to $x$.
If there exists a node $v$, the hole detection ratio from $p$ to whom is larger than a predefined threshold $\delta$, that is,\\
 \text{~~~~~~~~}$length\_pro(p, v)/dist\_euc(p, v)$ $>$ $\delta$, \text{~~~~~~~~}(1) \\
then $p$ is considered sitting on the boundary of a hole.

\subsection{Derivation of the threshold}
\begin{figure}[!htp]
\begin{center}
\includegraphics[width=5.0cm]{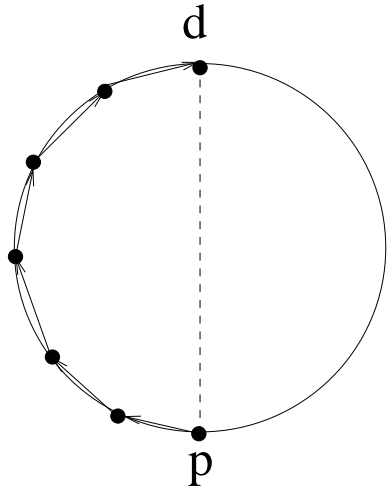}
\end{center}
\caption{Illustration of the shape of a hole: scenario 1}
\label{dis1fig}
\end{figure}

The value of $\delta$ has essential impact to the results of  hole detection. If the value of $\delta$ is too large, it introduces
false negatives. If  $\delta$ is too small, it causes false positives. We know that when a hole exists, there will be a detour path. So we attempt
to detect a hole by finding a detour path. In our approach, ``detour" path is defined as the routing path
between two nodes that is much longer than their Euclidean distance. In order to quantitatively represent
``much longer,"  we introduce a threshold $\delta$ for the ratio of routing length over Euclidean distance $length\_pro()/dist\_euc()$. To determine the value of $\delta$,
we first approximate the polygon by a circle (Fig.~\ref{dis1fig}) in which $\delta$ is $\pi/2=1.57$. But the circle cannot be a hole because no node in its circumference can be a local minimum node,
so we will investigate the cases $\delta>1.57$.

\begin{figure}[!htp]
\begin{center}
\includegraphics[width=5.0cm]{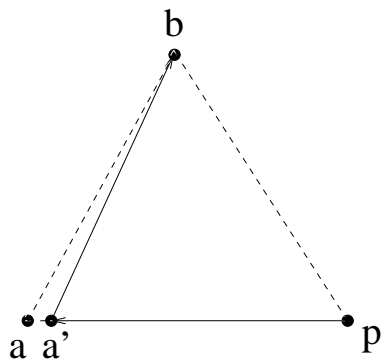}
\end{center}
\caption{Illustration of the shape of a hole: scenario 2}
\label{dis2fig}
\end{figure}

We then increase the value of $\delta$. Suppose that triangle $abp$ is a equilateral  triangle (Fig.~\ref{dis2fig}), the length of each edge is 1,
and the transmission range is slightly less than 1, such as 0.9. Then we move $a$ to $a^{'}$ and let both $a^{'}p$ and $a^{'}b$ be equal to the
transmission range. Then from $p$ to $b$, a path $p \rightarrow a' \rightarrow b$ exists and it is a slight detour path.
But the triangle is not a hole since none of the three nodes is a local minimum node. In this circumstance,
the value of $\delta$ is approximate equal to 2.

Then we increase the value of $\delta$ to the one that is slightly larger than 2.
By analysis and experiments, we found that $\delta$ =2.25 is a good choice for a small false positive and a small false negative
by experimental attempts.

\subsection{False Negative and False Positive of Hole Detection}
\begin{figure}[!htp]
\begin{center}
\includegraphics[width=5.0cm]{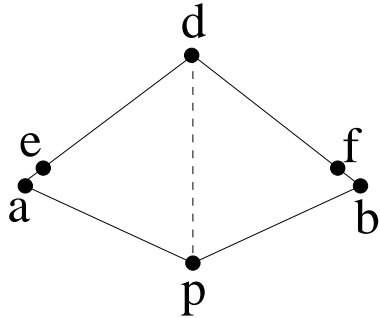}
\end{center}
\caption{Illustration of a false negative when detecting holes by HDDL}
\label{dis3fig}
\end{figure}

False negatives  and false positives may occur during hole detection.
Fig.~\ref{dis3fig} shows an example of false negative. In this figure, the transmission range is 0.9, $\mid pd \mid$=0.95, $\mid ad \mid$=$\mid bd \mid$=1,$\mid ed \mid$=$\mid df \mid$=0.9
and $\mid ap \mid$=$\mid pb \mid$=0.9. In this scenario, $P$ cannot talk to $d$ directly. Then $p$ is a local minimum node and the polygon $paedfb$ is a hole. However, if $p$ initiates
a probe message, the distance of the probe path is 1.9. And the Euclidean distance is 0.95. The ratio of the two distances is 2, which is less than 2.25. Consequently,
HDDL
does not consider that polygon $paedfb$ is a hole. False negative is introduced by a very special circumstance that the detour path is
between 2 to 2.25 long as the Euclidean distance, and the probe message initial node cannot talk to the destination directly.
In network environment, the possibility of false negative is low. Moreover,
false negative  will not affect the routing too much
because the detour is not too long, normally one hop longer than the Euclidean distance.

\begin{figure}[!htp]
\begin{center}
\includegraphics[width=7.0cm]{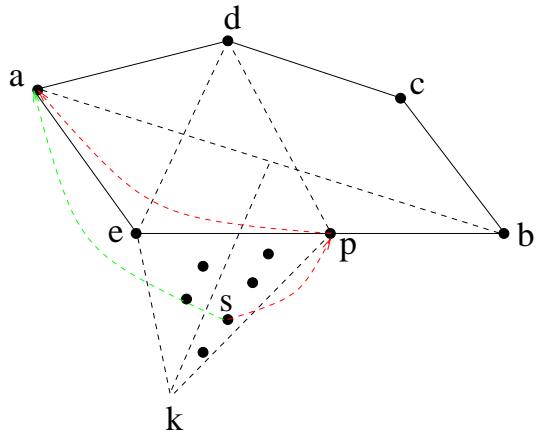}
\end{center}
\caption{Illustration of a false positive when detecting holes by HDDL}
\label{dis4fig}
\end{figure}

False positive can also occur in some special circumstances.
 For instance, in
Fig.~\ref{dis4fig},
$\mid pd \mid$=1, $\mid ed \mid$=0.95,
and $\mid ep \mid$=$\mid ea \mid$=$\mid ad \mid$=$\mid dc \mid$=$\mid cb \mid$=0.9. Once $p$ wants to talk to $d$, $p$ can find its neighbor $e$
that is closer to $d$. Hence, $p$ is not a local minimum node and then polygon $peadcb$ is not a hole.   
However, if $p$ initiates a probe message, HDDL
considers that the polygon is a hole. Furthermore, node $e$ and $p$ will advertise the hole information to an area $epk$.
The area to be announced the hole information is not big
because only a few nodes (two nodes in this example) announce the hole information.
Although the polygon is a fake hole detected by HDDL, the nodes in $epk$ will benefit from the hole information.
For example, if $s$ wants to send
a packet to $d$, $s$ will send the packet to $d$ directly instead of a detour from $p$.


False negatives and false positives  appear some time. However, their impact on HDDL algorithm is limited.

\subsection{Detection of Holes}
\begin{figure}[!htp]
\begin{center}
\includegraphics[width=7.0cm]{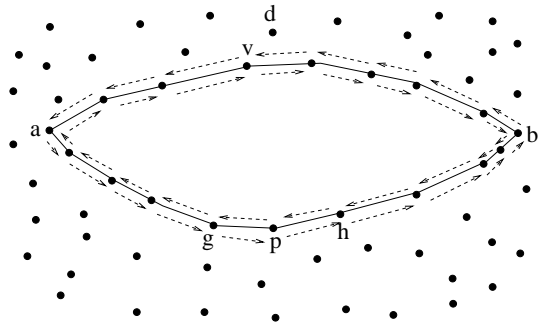}
\end{center}
\caption{Illustration of the approach of hole detection}
\label{hdarfig1}
\end{figure}

We derived that  $\delta$=2.25 is a good choice to detect
most holes that will block greedy forwarding.
Fig.~\ref{hdarfig1}  is an example for hole detection. Node $p$ initiates the hole probe message. $p$ collects the nodes'
locations while the message circulates the loop. If $p$ finds that there exists a node $v$,
satisfying $length\_pro(p, v)/dist\_euc(p, v)$ $>$ 2.25 , $p$ is considered to be sitting on a hole.

The hole that is detected is a polygon. Note that some nodes located on the polygon measure they are
located on the hole, but other nodes may not consider themselves on the hole. For instance,
in Fig.~\ref{hdarfig1}, nodes $g$, $p$ and $h$  consider themselves on the hole because there are nodes on the polygon's
boundary that satisfy the hole definition for nodes $g$, $p$ and $h$ (1). However,
nodes $a$ and $b$ at the hole polygon found by $p$ do not consider themselves on a hole because
there is no node on the hole satisfying condition (1) for nodes $a$ and $b$. In fact,
$a$ or $b$'s greedy
forwarding will not be blocked
by the polygon.

Any node located on the polygon may detect the hole repeatedly independently, thus a lot of overhead will be generated.
We design a mechanism to reduce the redundant probes for discovering the hole.
Once a node hears a probe message, it will not schedule a probe message although it has
not sent out its probe message yet.
In order to make each node know the location of every node on the polygon, the probe initiating node sends two probe
messages at the same time, one clockwise and one counter-clockwise. (shown in Fig.~\ref{hdarfig1}). In this way, each node on the polygon can obtain
all of the information of the polygon. Because the probe message is sent  in  both clockwise and counter-clockwise directions,
there will be two probe paths. We choose the longer as the length of the probe path to calculate the hole detection ratio.
We describe the probe message initiating algorithm in Fig.~\ref{initiatefig}.

\begin{figure}[hbt]
\baselineskip=10.2pt
{ \small
  \noindent\underline{\bf Probe\_msg\_initiating()} \\
  \text{~~~~}\textbf{if} does not receive a probe message \\
  \text{~~~~~~~~}\textbf Search whether it has an angle between two adjacent edges \\
  \text{~~~~~~~~}larger than 120 degrees; \\
  \text{~~~~~~~~}\textbf{if} true \\
  \text{~~~~~~~~~~~~}Initiate a probe message; \\
  \text{~~~~~~~~~~~~}Write its location to this message; \\
  \text{~~~~~~~~~~~~}Send it to its left node and right node of the angle.\\
 }

\protect\caption{Probe Message Initiating Algorithm}
\protect\label{initiatefig}
\end{figure}

The probe message receiving algorithm is described in Fig.~\ref{receivefig}. In this algorithm,
upon receiving a probe message, a node determines whether it is message initiation node. If it is,
the node will calculate the hole detection ratio when both probe massages come back.  If the node is not the
message initiation node, it will write its location to the message and forward the message.

\begin{figure}[hbt]
\baselineskip=10.2pt
{\small
  \noindent\underline{\bf Probe\_msg\_receiving()} \\
  \text{~~~~}Compare its location with the message initiator's location to \\
  \text{~~~~}determine whether it is the message initiator \\
  \text{~~~~}\textbf{if} it is the initiator \\
  \text{~~~~~~~~}Search whether the two probe message from  \\
  \text{~~~~~~~~}different directions both reached it; \\
  \text{~~~~~~~~}\textbf{if} true \\
  \text{~~~~~~~~~~~~}Calculate the hole detection ratio \\
  \text{~~~~~~~~}\textbf{else} \\
  \text{~~~~~~~~~~~~}Wait for the second probe message; \\
  \text{~~~~}\textbf{else} \\
  \text{~~~~~~~~}Write its location;\\
  \text{~~~~~~~~}Forward the message to its left or right neighbor\\
  \text{~~~~~~~~}according to the forwarding direction.\\
 }

\protect\caption{Probe Message Receiving Algorithm}
\protect\label{receivefig}
\end{figure}

The probe initiator must have an angle between two adjacent edges with respect to it that is larger than 120$^\circ$ ~\cite{fang31}.
However, such an angle is necessary but not a sufficient condition to determine if the initiator is a local minimum node.
In our algorithm, it does not matter whether the probe initiator is a local minimum node or not.
The objective of the hole probe message is to find a hole, but not to determine if
the probe initiator is a local minimum node.

\begin{figure}[!htp]
\begin{center}
\includegraphics[width=7.0cm]{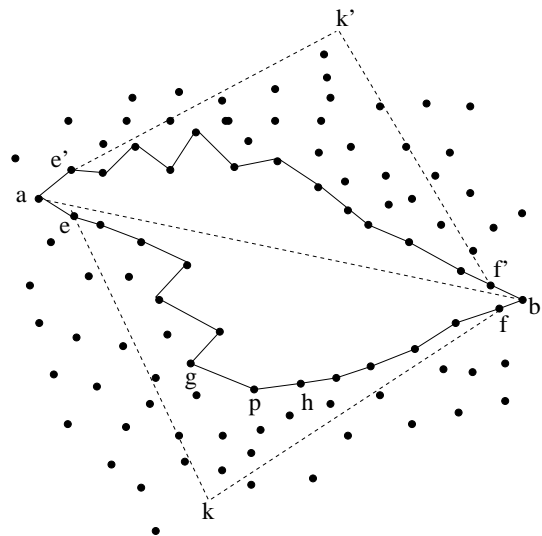}
\end{center}
\caption{Hole detection scenario: Example 1}
\label{hdarfig2}
\end{figure}

Most likely, a probe initiator that finds a hole  is
a local minimum node. For example, in Fig.~\ref{hdarfig1}, node $p$ initiates the  probe message and finds that it is
located on a hole. It is a local minimum node if it sends a packet to nodes in the vicinity of node $d$. However, it is not necessary for the probe initiator
to be a local minimum node. For instance, in Fig.~\ref{hdarfig2}, node $p$
initiates a hole probe message and detects a hole, but $p$ is not a local minimum node because either its neighbor $g$ or $h$ is closer to any destination node in the area in the opposite side of $ab$. The hole information
will be announced to the nodes in a certain area and these nodes will benefit from the hole announcement for future routing.

\begin{figure}[!htp]
\begin{center}
\includegraphics[width=7.0cm]{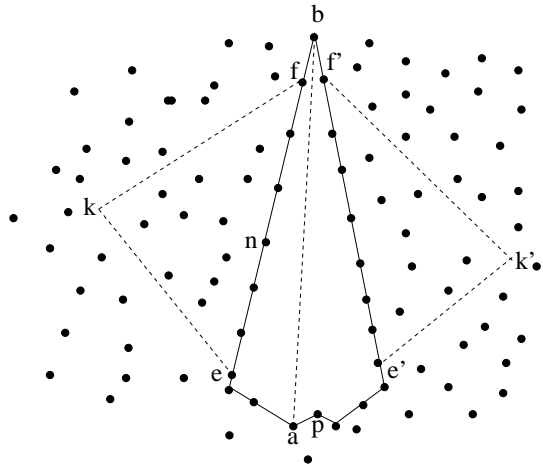}
\end{center}
\caption{Hole detection scenario: Example 2}
\label{hdarfig3}
\end{figure}

In Fig.~\ref{hdarfig3}, $p$ is a local minimum node. $p$ initiates a hole probe message but it cannot detect the hole
because the length of
the probe path from $p$ to any node on the polygon over their Euclidean distance is approximate to 1. However, the hole can be
detected by another node such as $n$ and the hole information will be announced to nodes in areas ($ekf$ and $e^{'}k^{'}f^{'}$) containing
the nodes which will benefit
from the hole information in future routing. This phenomenon indicates the scenario that a local minimum node $p$ cannot detect a hole.
This is because the polygon is long and narrow, and then the initiator's
routing will not be blocked by the polygon. So this detecting result has very minor effect on $p$'s forwarding.
Nevertheless, the hole will be detected by another node who suffers from the hole.

\subsection{Shape-Free Hole Representation}
A hole that is detected is a polygon. The representation of a polygon is a sequence of vertices.
However, in geographic routing, we do not have to care about all the nodes on the polygon because most of
them have a minor impact on determining the routing paths. What we are interested are the nodes that will block the
greedy forwarding.
In our model,
$p$ can calculate the two nodes whose Euclidean distance is most remote because  $p$ has already obtained all the nodes' locations on the polygon (Fig.~\ref{hdarfig2}). The segment connecting these
two nodes looks like a board that blocks greedy forwarding. For instance, segment $ab$ in Fig.~\ref{hdarfig2} is
a board that blocks greedy forwarding. Then the hole is represented as $<a,b>$.
No matter what the shape of the hole is, what we are concerned is
the segment connecting the two most remote nodes. The size of the hole may change due to node failure or the addition of new nodes.
In order to detect and represent the hole accurately, node $p$ needs to send the information about the vertices  lying on the polygon to nodes $a$ and $b$ for future detection of size changes of the hole.

The greedy forwarding is stuck by the board $<a,b>$ is because some potential  destination nodes
are hidden behind the board, and source nodes located in a certain area on the opposite
side of these hidden destinations are not aware of these destination nodes. In the basic routing approach, each
node uses greedy forwarding until it fails due to a local minimum node, where greedy
forwarding changes to perimeter forwarding. Thus the detour paths are generated. If the possible destination nodes
hidden behind the board can be determined in advance and be announced to the source nodes unaware of these destination nodes,
the lengths of the routing paths can be  reduced dramatically. We determine the possible destination area (shaded area)
as follows. Draw line $ar$ perpendicular to segment $ab$, where $r$ and $p$ are on the opposite sides of $ab$. Also draw line $bt$
perpendicular to line $ab$, where $t$ and $p$ are on  the opposite sides of $ab$. Then the area $rabt$ is the shaded area (Fig.~\ref{hdarfig4}).

\subsection{Hole Announcement}
\begin{figure}[!htp]
\begin{center}
\includegraphics[width=7.0cm]{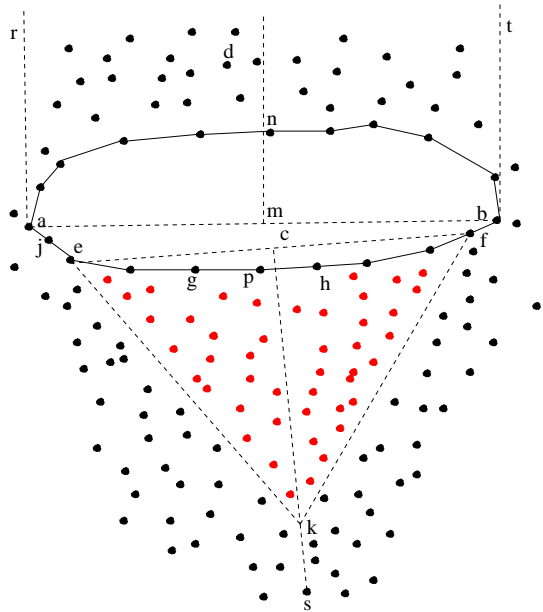}
\end{center}
\caption{A hole and the nodes in its vicinity area}
\label{hdarfig4}
\end{figure}

The nodes in area $rabt$ are the possible destination nodes for some source nodes.
We would like to figure out an area containing these source nodes that need to be announced
the hole information on the opposite side of $rabt$ (Fig.~\ref{hdarfig4}). The hole information can help
the nodes adaptively adjust the next forwarding hops to avoid detour routing paths.
In order to determine the hole announcement area, the announcement breadth and depth
need to be figured out. We first determine  two nodes $e$ and $f$. They are
the left and right nodes furthest away from each other at the same side as node $p$
of segment $ab$, and satisfy the hole
detection condition (1). Let $c$ be the midpoint of segment $ef$. Draw segment $ck$ perpendicular
to $ef$. Then triangle $efk$ is the area that should
be announced the hole information. Note that if the hole announcement area is larger,
more nodes will be benefited by the hole information and their future routing path will be shorter.
At the same time,  higher overhead will be introduced because more nodes need to be announced the hole information.
So we would like to find a good balance between the benefit to future routing paths and the overhead.

The optimal values of the hole information announcement size that can both shorten
the future routing path and reduce the overhead need to be found.

The announcement
breadth is selected as segment $ef$. So the announcement depth determines the size of the area.
Let $\mid ab \mid$ be $L$. Let $\mid ec \mid$ be $l$. Let $\angle cek$ be $\alpha$.
Then the objectives are to minimize the
number of nodes in the triangle  $ekf$ and minimize the length of the path from $s$ to $d$.
Assume that the nodes are distributed in the plane uniformly. So the number of the nodes in $\bigtriangleup$ekf
can be represented by the area of $\bigtriangleup$ekf:\\
\text{~~~~~~~~}$\frac{1}{2}$$\cdot$$2l$$\cdot$$l$$\tan\alpha$$=$$l^{2}$$\tan\alpha$,$\alpha\in\lbrack0,\frac{\pi}{2}\rbrack$ (2)\\
For node $s$, if it intends to send a packet to node $d$, the path includes the sub-paths $s \rightarrow k$,
$k \rightarrow e$, $e \rightarrow a$ and $a \rightarrow d$. The last two sub-paths are fixed,
but $s \rightarrow k$ and $k \rightarrow e$ depend on $\alpha$. Assume the length of $sc$ to be $h$.
We approximately represent the length of path $sk$ by $h-\mid kc\mid=h-l$$\tan\alpha$,
and the length of path $k \rightarrow e$ by $\mid ke\mid$ since the routing path generated by HDDL
will be along $ke$. Here  $\mid ke\mid$ is $\frac{l}{\cos\alpha}$. Hence from $s$ to $e$, the length of paths is:\\
\text{~~~~~~~~}$(h-l$$\tan\alpha)+\frac{l}{\cos\alpha}$     (3)\\
We want to find an $\alpha$ that can try to minimize both (2)and(3). Since (2) is quadratic to $l$ but (3) is linear to $l$,
so what we want to achieve is:\\
\text{~~~~~~~~}$argMin((2)*(3)^{2})$, that is:\\
\text{~~~~~~~~}$argMin(l^{2}\tan\alpha\cdot((h-l\tan\alpha)+\frac{l}{\cos\alpha})^{2})$\\
\text{~~~~~~~~}=$argMin(l^{2}h^{2}\tan\alpha+l^{4}\tan^{3}\alpha-2hl^{3}\tan^{2}\alpha\\
\text{~~~~~~~~}+\frac{l^{4}\tan\alpha}{\cos^{2}\alpha}+\frac{2hl^{3}\tan\alpha}{\cos\alpha}
-\frac{2l^{4}\tan^{2}\alpha}{\cos\alpha})$\\
Let $(l^{2}h^{2}\tan\alpha+l^{4}\tan^{3}\alpha-2hl^{3}\tan^{2}\alpha+\frac{l^{4}\tan\alpha}{\cos^{2}\alpha}+\frac{2hl^{3}\tan\alpha}{\cos\alpha}
-\frac{2l^{4}\tan^{2}\alpha}{\cos\alpha})$ be $g(\alpha)$. \\
Let $h=2l,2.5l,3l,3.5l,4l$ respectively, if there exists $\alpha_0$, $\alpha_1$ and $\alpha_2$ satisfying:\\
 \text{~~~~~~~~}$g(\alpha_1)'=0$, \\
 \text{~~~~~~~~}$g(\alpha_0)'<0$,\\
 \text{~~~~~~~~}and $g(\alpha_2)'>0$ \\
 where $\alpha_0$ is minor smaller than $\alpha_1$,$\alpha_2$ is minor larger than $\alpha_1$,
 then the expected $\alpha$ can be derived. Unfortunately, when  $g(\alpha_1)'=0$, $\alpha\notin\lbrack0,\frac{\pi}{2}\rbrack$.
 We substitute the series values of h to $g(\alpha)$, then
 achieve the minimal values of $g(\alpha)$ and their corresponding values of $\alpha$. The average value of $\alpha$ is 1.05.
 $\tan1.05=1.74$, so the depth of hole information announcement is:\\
 \text{~~~~~~~~}$1.74*l=1.74*\frac{L}{2}=0.87L$,\\
 Note  here $l$ is approximately represented by  $\frac{L}{2}$.

The nodes on arc $ef$ begin to advertise the hole information $<a,b>$ to their neighbors.
In order to avoid duplicate messages, once a node in
the area has received the hole information, it simply discards the duplicate.
After the  advertisement of the hole's information, each node in the area $efk$ is aware of the hole $<a,b>$. Consequently, these nodes
know that any possible destination node in area $rabt$ is hidden behind the hole. They should avoid  packets being forwarded towards the hole in future
routing (Fig.~\ref{hdarfig4}).


\section{Adaptive routing}
After the announcement, each node in the triangle $ekf$ knows that there is a hole $<a,b>$ that blocks greedy
forwarding to any destination node in area $rabt$. Thus the nodes in triangle $ekf$ can adaptively adjust routing
paths. In the network plane, once a node $s$ intends to send a packet with destination $d$, it first looks up its local
cache to see whether it has a hole information entry $<a,b>$. If there is no such entry in its cache, it just
uses GPSR. Otherwise if there is a hole information entry $<a,b>$, but $s$ and $d$ are located at the same side of segment
$ab$, $s$ just uses GPSR. If $s$ and $d$ are located on the opposite sides of $ab$ but $d$ is not in the shaded area $rabt$,
$s$ just uses GPSR. Otherwise $d$ lies in the area $rabt$. In this situation, $s$ considers $a$ or $b$ as its tentative
target. It writes $a$ or $b$ to the packet's header as a tentative target. In order to make $s$ determine which one
should be the tentative target, let $m$ be the midpoint of segment $ab$, $mn$ is perpendicular to segment $ab$ and
$n$ is on the opposite side of $ab$ relative to $s$. Then if $d$ is
located in area $ramn$, $s$ writes $a$ to the packet's head as its tentative target. If $d$ is located in area $nmbt$,
$s$ writes $b$ to the packet's head
as its tentative target. When the packet reaches $a$ or $b$, the tentative target will continue
sending the packet to the destination node $d$. In this approach, $a$ and $b$ are actually two landmarks. 
When a node forwards a packet to its next hop, it calls procedure $HDDL\_forwarding()$ given in Fig.~\ref{forwardfig}.

\begin{figure}[hbt]
\baselineskip=10.2pt
{\small
  \noindent\underline{\bf HDDL\_forwarding()} \\
  \text{~~~~}Look at the forwarding packet whether this node has a tentative \\
  \text{~~~~~~~~}target ${\cal T}$ \\
  \text{~~~~}\textbf{if} true \\
  \text{~~~~~~~~}Compare whether this node is ${\cal T}$;\\
  \text{~~~~~~~~}\textbf{if} true\\
  \text{~~~~~~~~~~~~}Remove ${\cal T}$ and forward the packet to next hop with its\\
  \text{~~~~~~~~~~~~~~~~}destination $d$;\\
  \text{~~~~~~~~}\textbf{else} \\
  \text{~~~~~~~~~~~~}Forward the packet to next hop with its destination ${\cal T}$;\\
  \text{~~~~}\textbf{else}\\
  \text{~~~~~~~~}Search local cache \\
  \text{~~~~~~~~}\textbf{if} an entry $<a,b>$ exists\\
  \text{~~~~~~~~~~~~}\textbf{if} this node and $d$ are at the same side of $ab$\\
  \text{~~~~~~~~~~~~~~~~}Use GPSR;\\
  \text{~~~~~~~~~~~~}\textbf{if} this node and $d$ are at opposite sides of $ab$\\
  \text{~~~~~~~~~~~~~~~~}\textbf{if} $d$ is in $ramn$\\
  \text{~~~~~~~~~~~~~~~~~~~~}Write $a$ as tentative target ${\cal T}$ to the packet;\\
  \text{~~~~~~~~~~~~~~~~~~~~}Forward packet to next hop with destination ${\cal T}$;\\
  \text{~~~~~~~~~~~~~~~~}\textbf{if} $d$ is in $nmbt$\\
  \text{~~~~~~~~~~~~~~~~~~~~}Write $b$ as tentative target ${\cal T}$ to the packet;\\
  \text{~~~~~~~~~~~~~~~~~~~~}Forward packet to next hop with destination ${\cal T}$;\\
  \text{~~~~~~~~~~~~~~~~}\textbf{else}\\
  \text{~~~~~~~~~~~~~~~~~~~~}Use GPSR;\\
  \text{~~~~~~~~}\textbf{else}\\
  \text{~~~~~~~~~~~~}Use GPSR.\\

 }

\protect\caption{Packet Forwarding Algorithm}
\protect\label{forwardfig}
\end{figure}

\section{Experimental Results}
We perform simulations using easim3D wireless network simulator,
which is used to simulate IEEE 802.11 radios and is typically used for location based routing algorithms. We
use a noiseless immobile radio network environment. In the
simulations, nodes with a transmission range of 20 meters
are deployed in an interest area of 400m$\times$400m.

We generate networks where the number of nodes varied
from 50 to 300. For any given number of nodes, 50 networks are generated randomly. In each network,
holes are generated automatically by the distribution of  nodes.

Our experiments include two parts.
The first part is to compare HDDL with GPSR and HAGR in terms of average length of paths and average hop of paths.

\begin{figure}
\begin{center}
\includegraphics[width=8.0cm]{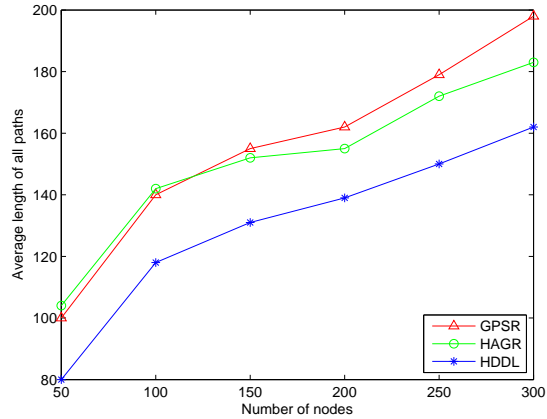}
\end{center}
\vspace{-0.1in}
\caption{Comparisons of average length for all paths}
\label{avepathfig}
\end{figure}

\begin{figure}
\begin{center}
\includegraphics[width=8.0cm]{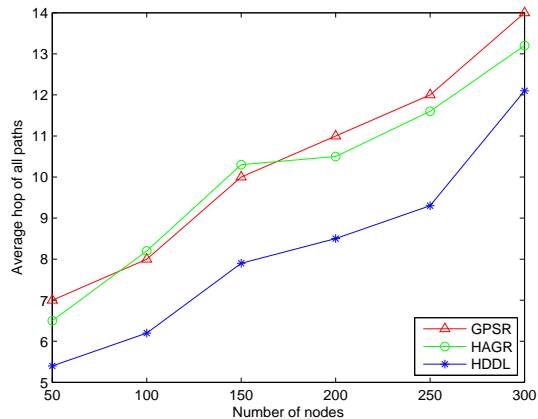}
\end{center}
\vspace{-0.1in}
\caption{ Comparisons of  average hop for all paths}
\label{hopsfig}
\end{figure}

Fig.~\ref{avepathfig} shows the average length of paths when the number of nodes changes from 50 to 300.
The average length in HDDL is 12.4\% shorter than that of GPSR and 11.8\% shorter than that of HAGR.
Fig.~\ref{hopsfig} shows the average number of hops in HDDL is 13.2\% less than that of GPSR and 12.7\% less than that of GPSR.

In Fig.~\ref{avepathfig} and Fig.~\ref{hopsfig}, the results report both the greedy path and
the path in the vicinity of holes. In this way,  HDDL's benefit for the paths near the holes is not highlighted. To demonstrate HDDL's impact for the paths around holes, we particular studied  the paths in the vicinity of holes.
we marked the paths that benefit from the hole information as ``hole paths" and recorded the pairs of source and destination nodes in our HDDL.
We  investigated the paths generated by GPSR and HAGR with the same pairs of source and
destination nodes. Then we compared the paths benefiting from hole information in HDDL
with the paths derived from GPSR and HAGR.

\begin{figure}
\begin{center}
\includegraphics[width=8.0cm]{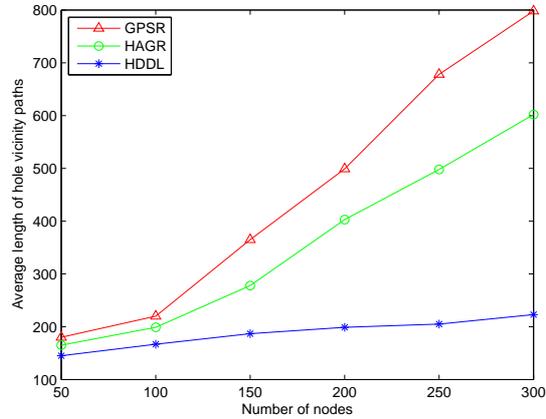}
\end{center}
\vspace{-0.1in}
\caption{ Comparisons of average length for hole vicinity paths}
\label{holepathfig}
\end{figure}

\begin{figure}
\begin{center}
\includegraphics[width=8.0cm]{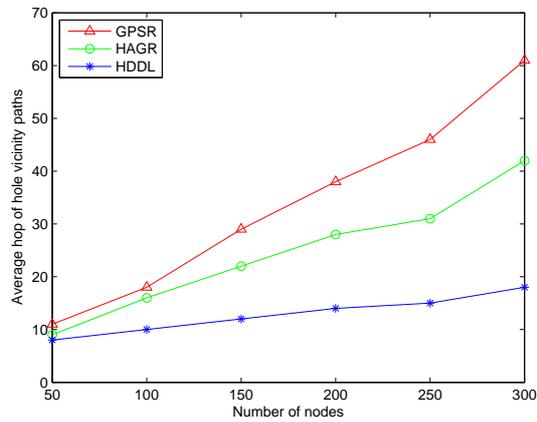}
\end{center}
\vspace{-0.1in}
\caption{ Comparisons of average hop for hole vicinity paths}
\label{holehopfig}
\end{figure}

The performance of HDDL,  GPSR and HAGR for hole paths are reported in Fig.~\ref{holepathfig} and  Fig.~\ref{holehopfig}.
HDDL has much shorter paths and fewer hops compared with GPSR.
The two figures indicate that HDDL reduces the long detour paths around holes significantly.

\begin{figure}
\begin{center}
\includegraphics[width=8.0cm]{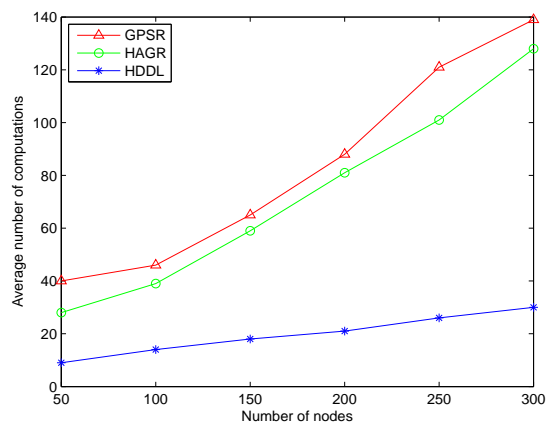}
\end{center}
\vspace{-0.1in}
\caption{Comparisons of computational complexity}
\label{complexityfig}
\end{figure}

The second part is to compare the computational
complexity of HARG and our algorithm HDDL.
We used the same networks as in part 1.
We selected $5\pi/6$ as the hole detection threshold and 60 meters as the diameter threshold.
In both HARG and HDDL, we investigated the number of computation times of hole detection.
In HDDL, the hole information is only calculated by a few nodes located on the hole and other
nodes are advertised the hole information. In HARG, a number of nodes have to perform calculation to determine
the existence of a hole. The numbers of calculations performed were reported to evaluate the computational complexity.
Fig.~\ref{complexityfig} illustrates that the computational complexity of HDDL is much less than that of HARG.

\newpage

\bibliographystyle{elsarticle-num}
\bibliography{egbib}
~~~\\
~~~\\







\end{document}